%% file: healthcareIoTSecurity.tex
\documentclass[10pt,conference,letterpaper]{IEEEtran}
\pdfoutput=1

\usepackage{url}
\usepackage{amsmath}
\usepackage{amssymb}
\usepackage{graphicx}
\usepackage{epstopdf}
\usepackage{caption}
\usepackage{enumitem}
\usepackage{algorithm2e}
\usepackage{color}
\usepackage[noadjust]{cite}
\newlist{myitemize}{itemize}{2}
\setlist[myitemize,1]{label=\textbullet,leftmargin=5mm}
\setlist[myitemize,2]{label=$\rightarrow$,leftmargin=6mm}
\setlist{parsep=0pt,listparindent=\parindent}
\makeatletter
\newcommand{\removelatexerror}{\let\@latex@error\@gobble}
\makeatother

\newcommand{\ignore}[1]{}
\newcommand{\dsse}{{\ensuremath{\sf DSSE}}}

\newcommand{\pseudo}{{\mathcal{F}}}
\newcommand{\hash}{{\mathcal{H}}}

\newcommand{\filter}{{\ensuremath{\sf BF}}}
\newcommand{\filteradd}{{\ensuremath{\sf BFAdd}}}
\newcommand{\filterverify}{{\ensuremath{\sf BFVerify}}}
\newcommand{\id}{{\ensuremath{\sf ID}}}

\newcommand{\hashtable}{{\ensuremath{\sf TBL}}}
\newcommand{\tblval}{{\ensuremath{\sf val}}}
\newcommand{\tblkey}{{\ensuremath{\sf key}}}
\newcommand{\counter}{{\ensuremath{\sf cnt}}}

\newcommand{\key}{{\mathcal{K}}}
\newcommand{\genkey}{{\ensuremath{\sf GenKey}}}
\newcommand{\addfile}{{\ensuremath{\sf AddFile}}}
\newcommand{\gentoken}{{\ensuremath{\sf GenToken}}}
\newcommand{\search}{{\ensuremath{\sf Search}}}
\newcommand{\sseverify}{{\ensuremath{\sf SSEVerify}}}

\newcommand{\keyset}{{\ensuremath{\sf K}}}
\newcommand{\state}{{\ensuremath{\sf state}}}
\newcommand{\token}{{\ensuremath{\sf token}}}
\newcommand{\result}{{\ensuremath{\sf rst}}}
\newcommand{\prev}{{\ensuremath{\sf prev}}}
\newcommand{\ind}{{\ensuremath{\sf Ind}}}

\newcommand{\se}{{\ensuremath{\sf SE}}}
\newcommand{\enc}{{\ensuremath{\sf Enc}}}
\newcommand{\dec}{{\ensuremath{\sf Dec}}}

\newcommand{\mac}{{\ensuremath{\sf Mac}}}
\newcommand{\macgen}{{\ensuremath{\sf GenMac}}}

\newcommand{\resultproof}{{\ensuremath{\sf prof}}}
\newcommand{\position}{{\ensuremath{\sf pos}}}
\newcommand{\digit}{{\ensuremath{\sf digit}}}

\newtheorem{definition}{Definition}

\begin{document}

\title{RSPP: A Reliable, Searchable and Privacy-Preserving e-Healthcare System for Cloud-Assisted Body Area Networks}
%\title{Dynamic Search Symmetric Encryption with Verifiability for Heath Data in Cloud-Assisted Body Area Network}

\author{\IEEEauthorblockN{Lei Yang}
\IEEEauthorblockA{Department of Electrical 
\\ Engineering and Computer Science\\
The University of Kansas, KS, 66045}
\and
\IEEEauthorblockN{Qingji Zheng}
\IEEEauthorblockA{
Bosch Research and Technology Center\\
Robert Bosch LLC\\
Pittsburgh, PA, 15222}
\and
\IEEEauthorblockN{Xinxin Fan}
\IEEEauthorblockA{
Bosch Research and Technology Center\\
Robert Bosch LLC\\
Pittsburgh, PA, 15222}}

% make the title area
\maketitle

\begin{abstract}
The integration of cloud computing and Internet of Things (IoT) is quickly becoming the key enabler for the digital transformation of the healthcare industry by offering comprehensive improvements in patient engagements, productivity and risk mitigation. This paradigm shift, while bringing numerous benefits and new opportunities to healthcare organizations, has raised a lot of security and privacy concerns. In this paper, we present a reliable, searchable and privacy-preserving e-healthcare system, which takes advantage of emerging cloud storage and IoT infrastructure and enables healthcare service providers (HSPs) to realize remote patient monitoring in a secure and regulatory compliant manner. Our system is built upon a novel dynamic searchable symmetric encryption scheme with forward privacy and delegated verifiability for periodically generated healthcare data. While the forward privacy is achieved by maintaining an increasing counter for each keyword at an IoT gateway, the data owner delegated verifiability comes from the combination of the Bloom filter and aggregate message authentication code. Moreover, our system is able to support multiple HSPs through either data owner assistance or delegation. The detailed security analysis as well as the extensive simulations on a large data set with millions of records demonstrate the practical efficiency of the proposed system for real world healthcare applications.   
\end{abstract}

\IEEEpeerreviewmaketitle

\input{./introduction/introduction.tex}
\input{./problemformulation/problemformulation.tex}

\input{./basic-construction/basicconstruction.tex}

\input{./full-construction/fullconstruction.tex}
\input{./security/security.tex}
\input{./performance/performance.tex}

\input{./relatedwork/relatedwork.tex}
\input{./conclusion/conclusion.tex}

\bibliographystyle{IEEEtran}
\bibliography{healthcareIoTSecurity}

% that's all folks
\end{document}

%% file: introduction/introduction.tex
\section{Introduction}
\label{intro}
In recent years, with the fast development of cloud computing and Internet of Things (IoT), the conventional healthcare industry is being reshaped to a more flexible and efficient paradigm of e-healthcare.  In a typical e-healthcare setting, a group of wearable and/or implantable devices (e.g., smart watches, bracelets, pacemakers, etc.), which forms a wireless body area network (BAN), gathers key vital signs (e.g., heart rate, blood pressure, temperature, pulse oxygen, etc.) from patients at home periodically. Those information is aggregated into a single file called personal health information (PHI) at an IoT gateway and then forwarded to a cloud server for storage. Third-party healthcare service providers (HSPs) can monitor patients' PHI and provide timely diagnosis and reactions by submitting on-demand queries to cloud storage. Although the increasing adoption of cloud computing and IoT services in healthcare industry helps reduce IT cost and improves patient outcomes, security and privacy of PHI are still major concerns as highlighted by the numerous reported data breaches due to malicious attacks, software bugs or accidental errors~\cite{databreach}. In particular, the healthcare regulations such as the Health Insurance Portability and Accountability Act (HIPAA) explicitly require that PHI be secured even as it migrates to the cloud.

While simply encrypting PHI before outsourcing it to the cloud can ensure the regulatory compliance of a healthcare system, it makes PHI utilization (e.g., query by third party HSPs) particularly challenging. Searchable encryption technology (see~\cite{song2000practical, curtmola2006searchable, boneh2004public} for pioneering work), which allows encrypted documents to be searched as is by augmenting them with an encrypted search index, provides a promising solution to addressing the aforementioned dilemma. An important line of research on searchable encryption is searchable symmetric encryption (SSE), which is considered more practical in terms of search efficiency for large datasets when compared to its public key-based counterpart. During the past decade, many provably secure SSE schemes~\cite{curtmola2006searchable,wang2010secure, cash2013highly,kamara2012dynamic, kamara2013parallel, naveed2014dynamic, DBLP:conf/ndss/CashJJJKRS14, hahn2014searchable,stefanov2014practical} have been proposed, which make trade-offs among security, search performance and storage overhead by exploring static-~\cite{curtmola2006searchable,wang2010secure,cash2013highly} and dynamic datasets~\cite{kamara2012dynamic,kamara2013parallel, naveed2014dynamic,DBLP:conf/ndss/CashJJJKRS14,hahn2014searchable,stefanov2014practical} as well as various data structures such as an inverted index~\cite{curtmola2006searchable,kamara2012dynamic}, a document-term matrix~\cite{kurosawa2013update}, a dictionary~\cite{DBLP:conf/ndss/CashJJJKRS14}, etc. 

We note that previous SSE schemes mainly focus on general search applications on encrypted database. The static SSE schemes that process static datasets and do not support subsequent updates are clearly not suitable for our e-healthcare applications. Moreover, most previous dynamic SSE schemes (except for \cite{hahn2014searchable}) work on a setting where a large static dataset is first processed and outsourced to the cloud storage, followed by a number of (infrequent) update operations, which is quite different from the e-healthcare applications where PHI files are created and uploaded to the cloud periodically at a fixed frequency (e.g., every 10 minutes). To prevent the cloud server from inferring sensitive information related to a patient (e.g., activity pattern, diet habit, etc.) based solely on observation of the stored encrypted indices, a dynamic SSE scheme with forward privacy\footnote{For a dynamic SSE scheme, forward privacy means that when a new keyword and file identifier pair is added, the cloud server does not know anything about this pair~\cite{stefanov2014practical}.} is highly desirable. In addition, for remotely monitoring patients' health status, HSPs should be able to perform search on PHIs encrypted by patients. Hence, our system should support a multi-user setting where the data owner and data user might be different. Last but not least, the reliability of an e-healthcare system is also critical and any incorrect or incomplete search results could lead to significant consequences, thereby highlighting the requirement for a verification mechanism to be deployed into the system. 

Motivated by the above observations, we present a reliable, searchable and privacy-preserving e-healthcare system for cloud assisted BAN in this paper. The proposed system is built upon a novel dynamic SSE scheme with forward privacy and delegated verifiability, which enables both patients and HSPs to conduct privacy-preserving search on the encrypted PHIs stored in the cloud and verify the correctness and completeness of retrieved search results simultaneously. Our contributions can be summarized as follows:
\begin{enumerate}
\item We proposed a dedicated and efficient dynamic SSE scheme for e-healthcare applications where PHIs are generated and stored in the cloud periodically. Our scheme is able to achieve a sub-linear search efficiency and forward privacy by maintaining an increasing counter for each keyword at an IoT gateway.
\item We presented an efficient mechanism that provides patient-controlled search capability for HSPs, thereby extending our system to a multi-user setting. This desired property is realized through a novel application of Bloom filter~\cite{bloom1970space} on the data owner (i.e., a patient) side.
\item We designed a lightweight delegated verification scheme based on a combination of Bloom filter, Message Authentication Codes (MACs) and aggregate MACs~\cite{katz2008aggregate}, which enables patients to delegate the capability of verifying the search results to HSPs.
\end{enumerate}
     
The rest of this paper is organized as follows. We state system model, and design goals in Section~\ref{problemformulation} and introduce the notations and preliminaries in Section ~\ref{sec:notation}. To make the exposition clear, we present the basic construction of our novel dynamic SSE with forward privacy in Section~\ref{sec:basic-construction}, and extend it to the full construction supporting the multi-user setting and verifiability in Section~\ref{sec:full-construction}. We show security analysis and performance evaluation in Section~\ref{security} and Section~\ref{performance}, respectively. Finally, we discuss the related work in Section~\ref{relatedwork} and conclude this work in Section~\ref{conclusion}.

%% file: problemformulation/problemformulation.tex
\section{Problem Formulation}
\label{problemformulation}

\subsection{System Model}
\begin{figure}[t]
	\begin{center}
		\includegraphics[width=9cm, height=4.5cm]{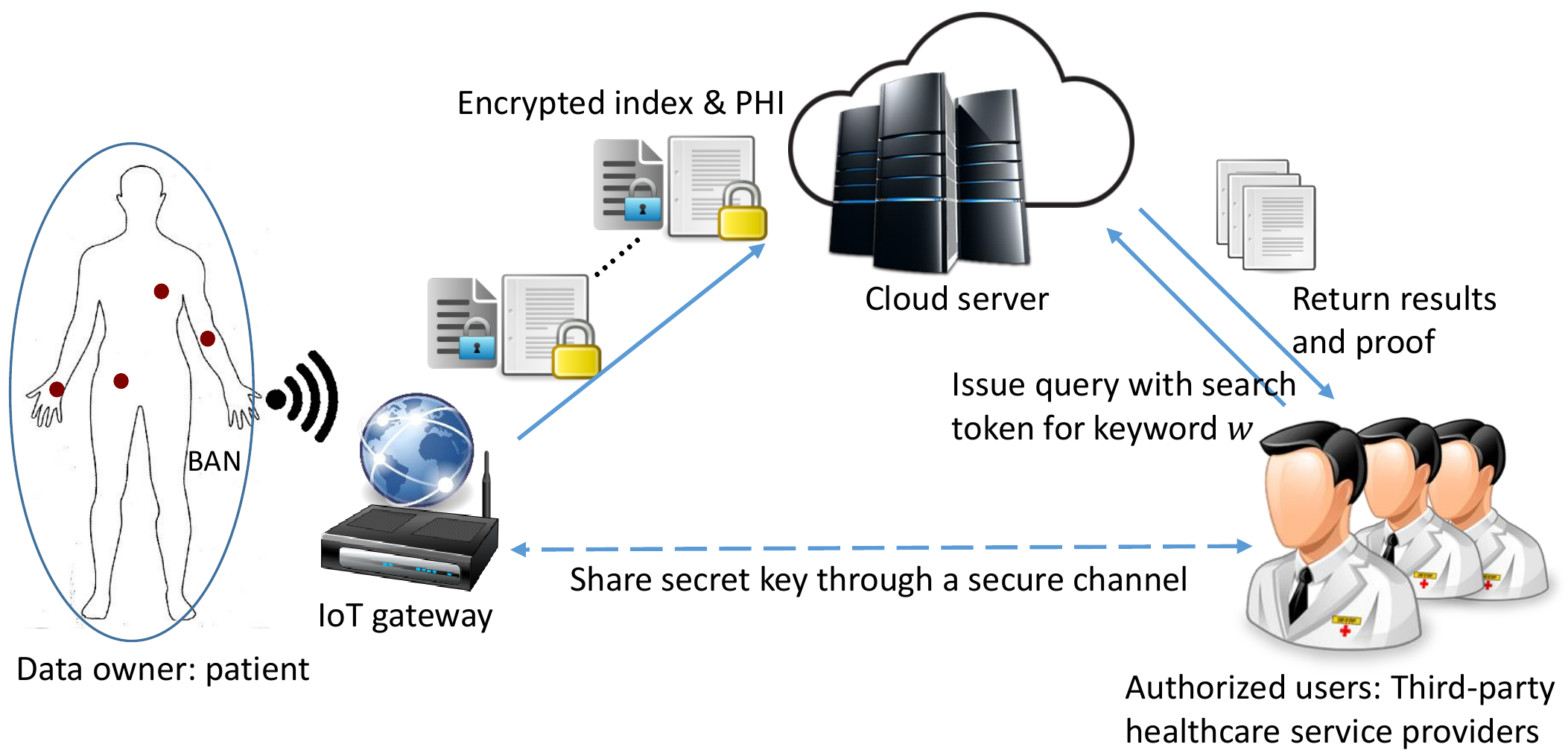}%[width=8.5cm, height=5cm]
	\end{center}
	\setlength{\abovecaptionskip}{-3pt}
	\setlength{\belowcaptionskip}{-10pt}
	\caption{The system model of a reliable, searchable and  privacy-preserving e-healthcare system.}%\protect\footnotemark
	\label{framework}
\end{figure}

The system model of our proposed reliable, searchable and privacy-preserving e-healthcare system involves four entities as shown in Fig.~\ref{framework}: a patient, an IoT gateway, a cloud server and several HSPs. The patient is the data owner whose health status is monitored by a group of wearable devices forming a BAN. The IoT gateway is the data aggregator which aggregates the periodically collected data into a single PHI file, extracts keywords, builds an encrypted index, and encrypts the PHI files. The encrypted index and PHI files are then sent to the cloud server for storage. Multiple HSPs act as the data users that provide healthcare services for the patient by querying and retrieving his/her encrypted PHIs from the cloud. We note that the e-healthcare system described above has the following unique properties with respect to data processing:
\begin{itemize}
	\item The PHI files are created by the IoT gateway and stored in the cloud \emph{periodically} (e.g., every 10 minutes).
	\item The PHI files are \emph{always added} into the cloud storage and file deletion or modification is not needed.
	\item The total number of unique keywords extracted from all the PHI files is not very large, due to the limited range of values for vital signs.    
\end{itemize}
%\footnotetext[1]{In basic construction, data users get search tokens from data owner, while in full construction, they generate search tokens by themselves.}

\subsection{Threat Model}
As assumed in most previous work on SSE~\cite{song2000practical,curtmola2006searchable,kamara2012dynamic,stefanov2014practical,hahn2014searchable}, the cloud server is generally ``honest-but-curious'', thereby faithfully performing the protocol but making inferences about the stored encrypted documents and data owner's private information. More specifically, in our e-healthcare system, the cloud server might try to infer whether a newly uploaded PHI file contains certain keyword or two PHI files contain the same keyword. Furthermore, the cloud server may also observe the queries submitted by HSPs (so-called search pattern) or the search results (so-called access pattern) to determine whether the same keyword is being searched. Additionally, considering the possibility of accidental system errors on the cloud server as well as the potential attacks from external adversaries, the cloud server might return incorrect or incomplete search results to data user. Finally, we assume that there is no collusion between data users and cloud server, or between users.

\subsection{Design Goals}
\label{design-goals}
In this work, we aim to design a reliable, searchable and privacy-preserving e-healthcare system which enables third-party HSPs to provide healthcare services for patients by searching on their encrypted PHIs incrementally stored on the cloud in a privacy-preserving and verifiable manner. The design goals of our system are as follows:
\begin{enumerate}
	\item \textbf{Search efficiency}. The search complexity on the cloud should be optimally sub-linear $O(k)$, where $k$ is the number of PHIs containing the queried keyword.
	\item \textbf{Forward privacy}. The cloud should not learn whether the newly stored PHIs contain some specific keywords.
	\item \textbf{Multi-user support}. HSPs should be able to perform patient-controlled search on behalf of a patient.
	\item \textbf{Verifiability}. HSPs should be able to verify the correctness and completeness of the search results.
\end{enumerate} 
Note that hiding search and access patterns in a general SSE setting can be achieved using the oblivious RAM (ORAM)~\cite{goldreich1996software}. However, ORAM-based schemes, while providing strong protection for privacy, incur significant computational and communication overhead for search. To ensure the practicality of our system, we did not consider employing the ORAM based approach to protect those patterns in this work.

\section{Notations, Preliminaries and Definition}
\label{sec:notation}

%In this section, we first present the necessary notations and primitives used in the rest of the paper, and briefly introduce the definition for dynamic searchable symmetric encryption. 

\subsection{Notations and Preliminaries}
\begingroup
\setlength{\thinmuskip}{0mu}
Let $e \leftarrow S$ denote selecting an element $e$ from a set $S$ uniformly at random, $\{0,1\}^n$ be the set of binary strings of length $n$, $\{0,1\}^*$ be the set of all finite length binary strings, and $||$ denote the concatenation of two strings. The data file $f$ is uniquely identified by the identity $\id(f)$ and contains a set of distinct keywords $W(f)=\{w_1, \ldots, w_l\}$. Let $\hashtable$ be a hash table storing key-value pairs $(\tblkey, \tblval)$ such that $\hashtable[\tblkey] = \tblval$,  $\hashtable[\tblkey] := \tblval$ denote assigning  $\tblval$ to $\tblkey$, and $\tblkey \in \hashtable$ denote that $\tblkey$ is an element of the key set in $\hashtable$.

Let $\pseudo_1:\{0,1\}^\lambda \times \{0,1\}^*$ $\rightarrow \{0,1\}^\lambda,$ $\pseudo_2: \{0,1\}^\lambda $ $ \times \{0,1\}^* $ $\rightarrow \{0,1\}^{2\lambda}, $ $\pseudo_3: \{0,1\}^\lambda \times \{0,1\}^* \rightarrow $ $\{0,1\}^{3\lambda}$ be three secure pseudorandom functions and $\hash: \{0,1\}^* \rightarrow \{0,1\}^{\lambda}$ be a secure hash function. Let $\se = (\se.\genkey, \se.\enc, \se.\dec)$ be a semantic secure symmetric encryption where $\se.\genkey$ is the key generation algorithm, $\se.\enc$ is the encryption algorithm and $\se.\dec$ is the decryption algorithm. 
Let $\mac = (\mac.\genkey, \mac.\macgen)$ be a secure message authentication code scheme, where $\mac.\genkey$ is the key generation algorithm, and $\mac.\macgen$ is the message authentication code generation algorithm.

\endgroup

Bloom filter is a space efficient data structure to represent a set $S$ and allow efficient membership query. A Bloom filter $\filter$ is an array of $m$-bit, which are set to 0 initially, and associated with $k$ independent universal hash functions $\hash_1, \ldots, \hash_k$, such that $\hash_i: \{0,1\}^* \rightarrow \{0, \ldots, m-1\}$. Given $e\in S$, the bits with respect to $\hash_i(e), 1\leq i \leq k,$ are set to 1. To query whether $e$ is an element of $S$ or not, one can check whether all bits with respect to $\hash_i(e), 1\leq i \leq k,$ are equal to 1. If not, $e \notin S$ for sure. Otherwise, $e\in S$ in a high probability due to the false positive rate. Suppose the outputs of all hash functions are in uniform random distribution and $n$ elements are hashed into the $\filter$, the false positive rate is $(1- e^{-kn/m})^k$. A $\filter$ usually associates with two algorithms:
\begin{itemize}
\item $\filter \leftarrow \filteradd(\filter, e):$ This algorithm hashes an element $e$ into the Bloom filter $\filter$.
\item $\{0,1\} \leftarrow \filterverify(\filter, e)$: This algorithm outputs 1 if $e$ is an element of $S$ where all elements were hashed into $\filter$ (with certain false positive rate); and 0 otherwise.
\end{itemize}

\subsection{Definition for Dynamic Symmetric Searchable Encryption}

Similar to the notation \cite{stefanov2014practical}, let $((c_{out}),(s_{out})) \leftarrow protocol ((c_{in}), (s_{in}))$ denote the protocol running between the data owner and the server, where the data owner takes as input $c_{in}$ and outputs $c_{out}$, and the server takes as input $s_{in}$ and outputs $s_{out}$.

\begin{definition}
The verifiable $\dsse$ scheme that supports streaming data consists of the following algorithms/protocols:
\begingroup
\setlength{\thinmuskip}{0mu}

\begin{itemize}[leftmargin=5mm]
\item $\keyset \leftarrow \genkey(1^\lambda)$: Given a security parameter $\lambda$, the data owner runs the algorithm to generate the secret key $\keyset$.

\item $((\state'_c), (\state'_s, C)) \leftarrow \addfile((\keyset, \state_c, f)$, $(\state_s))$: The data owner takes as inputs the secret key $\keyset$, current state information $\state_c$ and the file $f$ containing a set of keywords $W(f)$, and the server takes as input its current state information $\state_s$. The data owner runs this protocol to outsource $C$ (the encryption form of the file $f$) to the server and  updates its own state to $\state'_c$. The server also updates its own state to $\state'_s$. Initially, both $\state_c$ and $\state_s$ are empty.

\item $\token \leftarrow \gentoken(\keyset, \state_c, w):$ The data owner runs this algorithm to generate  search token $\token$, by taking as input $\keyset$,  $\state_c$ and $w$.

\item $(\result, \resultproof) \leftarrow \search(\state_s, \token)$: Given the search token $\token$, the server runs this algorithm to output the search result $\result$ consisting of a set of file identifiers. Moreover, the server generates the proof $\resultproof$ showing the correctness of the search result.

\item $\{0,1\} \leftarrow \sseverify(\keyset, \state_c, w, \result, \resultproof)$: The data owner (or authorized user) runs this algorithm to verify the correctness of the search result $\result$, given $\keyset$, $\state_c$, $w$, and $\resultproof$.
\end{itemize}
\endgroup
\end{definition}
Basically, the verifiable $\dsse$ scheme supporting streaming data aims to achieve the following security goals: forward privacy, verifiability and  confidentiality of outsourced data and queried keyword.

%% file: basic-construction/basicconstruction.tex
\section{Dynamic SSE Achieving Forward Privacy}
\label{sec:basic-construction}

\begin{figure*}[htp]
\removelatexerror
\begin{center}
\fbox{
	\begin{tabular}{p{0.95\textwidth}}
			
\begin{itemize}
\item $\underline{\keyset \leftarrow \genkey(1^\lambda)}$: Let  $\pseudo_1:\{0,1\}^\lambda \times \{0,1\}^*\rightarrow \{0,1\}^\lambda,\pseudo_2: \{0,1\}^\lambda  \times \{0,1\}^*  \rightarrow \{0,1\}^{2\lambda}$ be two pseudorandom functions, $\hash: \{0,1\}^* \rightarrow \{0,1\}^{\lambda}$ be a secure hash function and $\se$ be a secure symmetric key encryption. Given the security parameter $\lambda$, the data owner selects  $\key \leftarrow \{0,1\}^{\lambda}$, runs $\se.\genkey$ to get  $\key_{\se}$, and sets $\keyset= (\key_{\se}, \key)$. 

\item $\underline{((\state'_c), (\state'_s, C)) \leftarrow \addfile((\keyset, \state_c, f),(\state_s))}$: Suppose that the identifier of file $f$ is $\id(f)$ and the set of keywords extracted from $f$ is $W(f)= \{w_1, \ldots, w_l\}$. Note that when the system was initialized, $\state_c = \hashtable_c = \emptyset$ and $\state_s = \hashtable_s = \emptyset$ where $\hashtable_c$ and $\hashtable_c$ are hash tables. The protocol proceeds as follows:

{\bf{The data owner:}}

\begin{algorithm}[H]
Let $\ind$ be an empty set, and run $C \leftarrow \se.\enc(\key_{\se}, f)$ for  file $f$ \linebreak
\For{each keyword $w \in W(f)$ }
						{
Let $\key_{\prev} =  0^{\lambda}$, $\counter = 1$ and $\counter_{\prev} = 0$ \linebreak 
							\If{$w \in \hashtable_c$}{
Retrieve $\counter$ from $\hashtable_c$ with respect to $w$ \linebreak
Let $\key_{\prev}= \pseudo_1(\key, \hash(w||\counter))$, $\counter_{\prev} = \counter$ and  $\counter = \counter+1$	}
Compute $\key_{\counter} \leftarrow \pseudo_1(\key, \hash(w||\counter))$ \linebreak
Compute $\tau_{\counter} = \pseudo_1(\key, w||\counter)$ and  $\mu_{\counter} = \langle \pseudo_1(\key, w||\counter_{\prev})|| \key_{\prev} \rangle \bigoplus \pseudo_2(\key_{\counter}, \tau_{\counter})$  \linebreak
Let $\hashtable_c[w]:= \counter$   and $\ind = \ind \bigcup \{(\tau_{\counter}, \mu_{\counter})\}$ }
Send $(C, \id(f), \ind)$ to the server and let $\state'_c = \hashtable_c$
				\end{algorithm}

{\bf{The server:}}

Upon receiving $(C, \id(f), \ind)$ from the data owner, the server proceeds as follows:

\begin{algorithm}[H]
\For{each  $(\tau, \mu) \in \ind$ }
						{
						Let $\hashtable_s[\tau]:=\mu||\id(f)$
					}
Store $C$ locally and set $\state'_s = \hashtable_s$
\end{algorithm}

\item $\underline{\token \leftarrow \gentoken(\keyset, \state_c, w)}$:  Given the keyword $w$ to be queried, the data owner generates the search token as follows: (i) Retrieve $\counter$ from $\state_c$ with respect to $w$, (ii) Compute $\key_\counter = \pseudo_1(\key, \hash(w||\counter))$ and (iii) Let $\token = (\pseudo_1(\key, w||\counter), \key_{\counter})$, which will be sent to the server.

\item $\underline{\result \leftarrow \search(\state_s, \token)}$: Given $\token = (\pseudo_1(\key, w||\counter), \key_{\counter})$, the server conducts the search by letting $\result$ be an empty set,  $\tau' = \pseudo_1(\key, w||\counter)$, $\key' =  \key_{\counter} $, and running the following algorithm:

\begin{algorithm}[H]

\While{$\key' \neq 0^{\lambda}$ }{
Retrieve $\mu||\id(f)$ from $\hashtable_s$ with respect to $\tau'$ and let $\result = \result \bigcup \{\id(f)\}$ \linebreak
Let $\tau'||\key' = \mu \bigoplus \pseudo_2(\key', \tau')$ (which results in $\pseudo_1(\key, w||(i-1)) || \key_{i-1}$ if the current counter is $i$)
}
Return $\result$ as the search result
\end{algorithm}

\end{itemize}
		\end{tabular}
	}
\end{center}
	\setlength{\abovecaptionskip}{-3pt}
	\setlength{\belowcaptionskip}{-10pt}
\caption{The $\dsse$ construction achieving forward privacy. Note that downloaded encrypted files can be decrypted with $\key_{\se}$.}
\label{basic-construction}
\end{figure*}

For the sake of simplicity, we first present the $\dsse$ construction achieving forward privacy, and leave the full-fledged $\dsse$ design to the next section. 

\subsection{Design Rational}
Informally, forward privacy in $\dsse$ demands that when adding a new file, the server should not learn whether the newly added file contains certain keyword that has been queried before or not, unless the keyword is queried again. Therefore, it is sufficient to achieve forward privacy if any keyword in the newly added file will not be linked to any encrypted keywords stored in the server.

Instead of using computationally heavy 
cryptographic primitives (e.g., ORAM), in this paper we exploit the combination of locally stored state information and chaining technique in a subtle way, and utilize the lightweight cryptographic primitives to achieve forward privacy, which is explained as follows.

The data owner associates to each keyword a counter, indicating the number of outsourced encrypted files having the keyword so far. That is, the data owner locally maintains the state information (i.e., pairs of keyword and counter). Suppose the counter associated to keyword $w$ is $\counter$, the index with respect to $w$, stored in the server, is a collection of tuples $\{
(\tau_1, \id(f_1) ), \ldots, (\tau_{\counter},\id(f_\counter))
\}$
\ignore{
\vspace{-0.2cm}
\begin{eqnarray*}
\tau_1,&& \id(f_1) \\
\tau_2, &&\id(f_2) \\
& \ldots & \\ 
%\end{eqnarray*}
%\begin{eqnarray*}
%\tau_{\counter-1} &=&\pseudo_1(\key, w||\counter-1),~~ \id(f_{\counter-1}) \\
\tau_{\counter}, && \id(f_{\counter})
\end{eqnarray*}
}
where $\tau_i = \pseudo_1(\key, w||i), 1\leq i \leq \counter, \pseudo_1$ is a secure pseudorandom function, $\key$ is a private key and $f_1, \ldots, f_{\counter}$ are files having keyword $w$. When adding a new file $f$ containing the keyword $w$, the data owner sends to the server the following tuple
\vspace{-0.25cm} 
$$(\tau_{\counter+1}, ~\id(f))$$
where $\tau_{\counter+1}=\pseudo_1(\key, w||\counter+1)$. Thanks to $\pseudo_1$,  without knowing $\key$ the server cannot know whether $\tau_{\counter+1}$ is generated from the same keyword as that of $\tau_{i}, 1 \leq i\leq \counter$. Note that the data owner does not need to maintain all previous states for each keyword because file deletion is not needed in healthcare. 

While binding counter to a keyword can break the correlation of two identical keywords, it raises another challenge: given one search token generated from the keyword and the counter, the server can only retrieve one single file identifier. That is, to retrieve all file identifiers having the specific keyword, the data owner has to enumerate all previous counters and  generate search tokens, which is rather costly in term of bandwidth for search.
%While maintaining state information locally can  achieve forward privacy, it raises another challenge: given one search token generated from the keyword and the counter, the server can only retrieve one single file identifier. That is, to retrieve all file identifiers having the specific keyword, the data owner has to enumerate all previous counters and  generate search tokens, which is rather costly in term of bandwidth for search.

\begingroup
\setlength{\thinmuskip}{0mu}
To mitigate this disadvantage, we use the following chaining technique, which implicitly links the tuples corresponding to the same keyword together (let $\tau_i = \pseudo_1(\key, w||i), 0\leq i \leq \counter$): 
\endgroup
\begin{eqnarray*}
\tau_1,  &
\langle \tau_0 || 0^{\lambda} \rangle \bigoplus \pseudo_2(\key_{1}, \tau_{1}), & \id(f_1) \\
\tau_2, & \langle \tau_{1} || \key_{1} \rangle \bigoplus \pseudo_2(\key_{2}, \tau_{2}), &\id(f_2) \\
& \ldots &  \\
\tau_\counter, &\langle \tau_{\counter-1} || \key_{\counter-1} \rangle \bigoplus \pseudo_2(\key_\counter, \tau_\counter),& \id(f_{\counter})
\end{eqnarray*}
\noindent where $\pseudo_2$ is another secure pseudorandom function and $\key_{i}, 1\leq i\leq \counter, $ is a  random key derived from the counter $i$. Obviously, without knowing $\key_i, i \geq \counter$, the server cannot correlate $\tau_{\counter}$ with $\tau_j, j<\counter$,
even though they might be generated from the same keyword (but different counter). On the other hand, given $\tau_{\counter}$ and $\key_{\counter}$, the server is able to obtain $\id(f_\counter)$ and recover $\tau_{\counter-1}$ and $ \key_{\counter-1}$ by computing
\begin{eqnarray*}
\langle \tau_{\counter-1} || \key_{\counter-1} \rangle \bigoplus \pseudo_2(\key_{\counter}, \tau_\counter) \bigoplus \pseudo_2(\key_\counter, \tau_\counter). 
\end{eqnarray*}
The server then obtains all file identifiers by iterating such process until that the key is $\lambda$-bit of zero. 

\subsection{Construction}
We show the construction in Fig.~\ref{basic-construction}. Here the random key $\key_{\counter}$ for keyword $w$ is generated by applying the pseudorandom function such that $\key_{\counter} = \pseudo_1(\key, \hash(w||\counter))$. In addition, the data owner stores the state information (i.e., pairs of $(w, \counter)$) in the hash table $\hashtable_c$, which maps keyword $w$ to the counter $\counter$.
%each entry records two elements: keyword $w$ and the counter $\counter$.
On the other hand, the server also stores the state information (i.e., the encrypted index) in the hash table $\hashtable_{s}$. We can see that given the keyword $w$, the search complexity is linear to the number of files containing $w$,  which is sublinear to the number of outsourced encrypted files.

\noindent{\bf Optimization I: Speed up search operation.} Note that the server might be able to speed up the search further: given $\token = (\tau_\counter, \key_{\counter})$ where $\tau_\counter= \pseudo_1(\key, w||\counter)$, the server can update its state information by setting $\hashtable_s[\tau_{\counter}] = \perp\| \result$, where $\perp$ is a stop sign and $\result$ is the search result with respect to $\token$. By doing this, %the server can  conduct the search without repeating the iterations especially when the same keyword is repeatedly queried. 
the server not only accelerates the search without repeating the iterations, but also saves the storage by storing file identifiers only.

%% file: full-construction/fullconstruction.tex
\section{Full-fledged DSSE Construction}
\label{sec:full-construction}
In this section, we present the full-fledged $\dsse$. In contrast to the $\dsse$ presented above, the full-fledged $\dsse$ not only achieves forward privacy, but also supports search capability enforcement and delegated verifiability, where the former allows the data owner (i.e., patients) to enforce controlled search capability, and the latter enables authorized data users (i.e., HSPs) to verify the correctness of the search result. 

\begin{figure*}[htp]
\removelatexerror
\begin{center}
\fbox{
	\begin{tabular}{p{0.95\textwidth}}
			
\begin{itemize}

\item $\underline{\keyset \leftarrow \genkey(1^\lambda)}$: Let $\pseudo_1:\{0,1\}^\lambda \times \{0,1\}^*\rightarrow \{0,1\}^\lambda,{\color{red} \pseudo_3: \{0,1\}^\lambda  \times \{0,1\}^*  \rightarrow \{0,1\}^{3\lambda}}$ be two pseudorandom functions, $\hash: \{0,1\}^* \rightarrow \{0,1\}^{2\lambda}$ be a secure hash function, $\se$ be a secure symmetric key encryption, {\color{red}  $\mac$ be a secure message authentication code}. Given the security parameter $\lambda$, the data owner selects  $\key \leftarrow \{0,1\}^{\lambda}$, runs $\se.\genkey$ to get  $\key_{\se}$, {\color{red}  runs $\mac.\genkey$ to get  $\key_{\mac}$}, and sets $\keyset= (\key, \key_{\se}, {\color{red} \key_{\mac} })$. 

\item $\underline{((\state'_c), (\state'_s, C)) \leftarrow \addfile((\keyset, \state_c, f),(\state_s)):}$ Suppose that the identifier of file $f$ is $\id(f)$ and the set of keywords extracted from $f$ is $W(f)= \{w_1, \ldots, w_l\}$. Note that when the system was initialized, $\state_c = (\hashtable_c = \emptyset, {\color{red} \filter_c = \emptyset})$ and  $\state_s = (\hashtable_s = \emptyset, {\color{red} \filter_s=\emptyset})$ where $\hashtable_c$ and $\hashtable_s$ are two hash tables, and $\filter_c$ and $\filter_s$ are two Bloom filters. The protocol proceeds as follows:

{\bf{The data owner:}}

\begin{algorithm}[H]
Let $\ind$ be an empty set, and run $C \leftarrow \se.\enc(\key_{\se}, f)$ for  file $f$ \linebreak
\For{each keyword $w \in W(f)$ }
{
Let $\key_{\prev} =  0^{\lambda}$,  $\counter_{\prev} = 0$, $\counter = 1$, {\color{red}$\gamma_{\prev} = 0^{\lambda}$ ~~~~~($\gamma_{\prev}$ is an aggregate MAC)} \linebreak 
	\If{$w \in \hashtable_c$}{
Retrieve $(\counter, {\color{red} \gamma_{\counter}})$ from $\hashtable_c$ with respect to $w$ \linebreak
Let $\key_{\prev}= \pseudo_1(\key, \hash(w||\counter))$, $\counter_{\prev} = \counter$,  ${\color{red}\gamma_{\prev} = \gamma_{\counter}}$, and $\counter = \counter+1$	}
Compute $\key_{\counter} \leftarrow \pseudo_1(\key, \hash(w||\counter))$, 
 ${\color{red} \gamma_{\counter} = \gamma_{\prev} \bigoplus \mac.\macgen(\key_{\mac}, C||w)}$ (The output of $\mac.\macgen$ is $\lambda$-bit length)\linebreak
Compute $\tau_{\counter} = \pseudo_1(\key, w||\counter)$,  $\mu_{\counter} = \langle \pseudo_1(\key, w||\counter_{\prev})|| \key_{\prev}||{\color{red}\gamma_{\counter}} \rangle \bigoplus {\color{red}\pseudo_3(\key_{\counter}, \tau_{\counter}})$  \linebreak
Compute ${\color{red} \filter_c \leftarrow \filteradd(\filter_c, \tau_{\counter})}$ \linebreak
Let $\hashtable_c[w]:= (\counter,  {\color{red} \gamma_{\counter}})$, $\ind = \ind \bigcup \{(\tau_{\counter}, \mu_{\counter})\}$ } 
{\color{red} Generate the MAC $\sigma \leftarrow \mac.\macgen(\key_{\mac}, \filter_c || T)$ where $T$ is the current time stamp} \linebreak
Send $(C, \id(f),  \ind, {\color{red}\sigma, T})$ to the server and let $\state'_c = (\hashtable_c, {\color{red} \filter_c})$
\end{algorithm}

{\bf{The server:}}

Upon receiving $(C, \id(f), \ind, \sigma, T)$ from the data owner, the server proceeds as follows:

\begin{algorithm}[H]
\For{each  $(\tau, \mu) \in \ind$ }
{
Let $\hashtable_s[\tau]:=\mu|| \id(f)$ and  {\color{red} $\filter_s \leftarrow \filteradd(\filter_s, \tau)$}
}
Store $C$ locally and set $\state'_s = (\hashtable_s, {\color{red} \filter_s, \sigma, T})$
\end{algorithm}

\end{itemize}

\noindent{\bf Suppose the data owner generated $r \leftarrow \se.\genkey$ and securely shared $r$ with authorized users and the server.}

\begin{itemize}
\item $\underline{\token \leftarrow \gentoken(\keyset, \filter_c, {\color{red} r}, w)}$: The data owner generates the search token as follows: (i)
Retrieve $(\counter, \gamma_{\counter})$ from $\hashtable_c$ with respect to $w$; (ii)
Compute $\key_{\counter} = \pseudo_1(\key, \hash(w||\counter))$; and (iii)
Let $\token = {\color{red} \se.\enc(r, \pseudo_1(\key, w||\counter)||\key_{\counter})}$, which will be sent to the server.

\item $\underline{(\result, \resultproof) \leftarrow \search(\state_s, {\color{red} r}, \token)}$:
The server runs ${\color{red} \se.\dec(r, \token)}$ to get $(\pseudo_1(\key, w||\counter) || \key_{\counter}$,
 and conducts the search by retrieving $\mu_{\counter} || \id(f)$ from $\hashtable_s$ with respect to  $\tau' = \pseudo_1(\key, w||\counter)$, computing $\mu_{\counter} \bigoplus \pseudo_3(\key_{\counter}, \tau')$ to {\color{red} get $\gamma_{\counter}$}, letting $\key' =\key_{\counter} $, ${\color{red} \resultproof = (\sigma, T, \filter_s, \gamma_{\counter}})$, $\result = \emptyset$, and

\begin{algorithm}[H]
\While{$\key' \neq 0^{\lambda}$ }{
Retrieve $\mu||\id(f)$ from $\hashtable_s$ with respect to $\tau'$, and let $\result = \result \bigcup \{\id(f)\}$ \linebreak
Let $\tau'||\key'||\gamma' =  \mu \bigoplus \pseudo_3(\key', \tau')$  (which results in $\pseudo_1(\key, w||(i-1)) || \key_{i-1}||\gamma_{i-1}$ if the current counter is $i$)
}
Return $\result$ as the search result and {\color{red} $\resultproof$ as the proof}
\end{algorithm}

{\color{red}
\item $\underline{\sseverify(K, w, \counter, \result, \resultproof)}$: Given $\resultproof = (\sigma, T, \filter_s, \gamma_{\counter})$, the data owner check whether the size of $\result$ is  equal to the counter $\counter$ or not. If not, then return 0 and abort. Otherwise, the verification proceeds as follows:
\begin{itemize}
\item Given $\id(f_i) \in \result, 1\leq i \leq \counter$, fetch encrypted data files $C_1, \ldots, C_{\counter}$ from the server.
\item If both equations hold, then output 1; otherwise output 0 (The data owner might not check Eq.(2) because of knowing correct $\counter$):
\begin{eqnarray*}
\bigoplus_{i=1}^{\counter}
 \mac.\macgen(\key_{\mac}, C_i||w)  \overset{?}{=} \gamma_{\counter} ~(1)~~~~~~~~~~~~~~~
 \mac.\macgen(\key_{\mac}, \filter_s||T) \overset{?}{=} \sigma ~(2)
\end{eqnarray*}
\end{itemize}
}

\end{itemize}
\end{tabular}
}
\end{center}
\caption{The full-fledged $\dsse$ construction achieving forward privacy, search capability enforcement and delegated verifiability. Note that the downloaded encrypted files can be decrypted with $\key_{\se}$.}
\label{full-construction}
\end{figure*}

\subsection{High Level Idea}

\noindent{\bf Search Capability Enforcement}. In order to enforce search capability, we need to resolve two questions: (i) how to grant authorized data users with search capability; (ii) how to revoke authorized data user's privilege if necessary. Furthermore, we require that the approach should be efficient without extensive interaction between the data owner and authorized data users. 

Granting search capability requires the data owner to distribute the secret key (i.e., $\key_{\se}$ and $\key$) and state information (i.e., the counter for each keyword) to authorized data users efficiently and securely. While secret key distribution can be done efficiently through a one-time off-line setup, state information distribution might be costly because authentication (between the data owner and the authorized user) is needed when authorized data users fetch the fresh state information, which is frequently updated. Note that making the data owner's state information public (even if encrypted) will harm the forward privacy because the server can infer which keyword (or encrypted keyword) was contained in the newly added file.

To address the above issue, we adopt the ``document-and-guess'' approach: The server maintains a Bloom filter $\filter_s$, and puts each received encrypted keyword $\pseudo_1(\key, w||\counter)$ into the Bloom filter $\filter_s$, and the authorized user, having the secret key already and fetching $\filter_s$ from the server, can guess the latest counter value by enumerating $(1,\ldots, \counter,\counter+1)$ such that $\pseudo_1(\key, w||\counter)$ is an element hashed to $\filter_s$ but $\pseudo_1(\key, w||\counter+1)$ not (suppose the false positive rate of $\filter_s$ is extremely low, e.g., $2^{-30}$ in our experiments).

%Note that for the state information $(\counter, \key_{\counter})$, $\key_{\counter}$ can be generated from $\counter$ with some pseudorandom function (suppose keys for pesudorandom functions have been secure distributed), so we focus on how to distribute $\counter$ for each keyword efficiently  (i.e., without extensive interaction). Our basic idea is to let the server maintain a Bloom filter $\filter$, and puts each received encryption form of keyword $\pseudo_1(\key, w||\counter)$ into the Bloom filter $\filter$. By doing this, when authorized data user wants to search for keyword $w$, it obtains $\filter$ from the server and then guesses the latest counter value by enumerating $(1,\ldots, \counter,\counter+1)$ such that $\pseudo_1(\key, w||\counter)$ is an element but $\pseudo_1(\key, w||\counter+1)$ not \footnote{Actually, the authorized data user can use the binary search for search $\counter$ by assuming $\counter \in [1, upper_bound]$ where $upper_bound$ can be some large enough value such as $2^{40}$}. We can see that without proper pseudorandom functions' key, no one can use $\filter$ to infer any state information (i.e., $\counter$).

On the other hand, in order to allow the data owner to revoke authorized users' search capability, we use the group key idea: The data owner generates a symmetric key $r$, which is securely shared with the server and all authorized users, such that the search token of keyword $w$ generated by authorized users should be $\se.\enc(r, \pseudo_1(\key, w||\counter)|| \key_{\counter})$ and the server can recover  $(\pseudo_1(\key, w||\counter), \key_{\counter})$ with the stored $r$ via $\se.\dec$, where $\se$ is a secure symmetric encryption. When an authorized data user was revoked, the data owner only needs to update the group key $r$ to $r'$ and the revoked user cannot generate valid search token without knowing $r'$.

\vspace{0.5em}
\noindent{\bf Delegated Verifiability}. The purpose of delegated verifiability is to allow authorized users (including the data owner) to verify that (i) correctness and completeness of search result, meaning the search result correctly consists of all file identifiers; and (ii) the integrity of the retrieved data files.

First, authorized users can leverage the counter value (if existing) to check whether the server returned the correct number of file identifiers because the counter value indicates the number of files having the specific keyword. Hence, in order to assure that authorized users get correct counter value (which is guessed from $\filter_s$), we need to enable the data user to verify that the cloud faithfully inserts the keywords into Bloom filer. To do so, the data owner also maintains a Bloom filter $\filter_c$, which is built from $\pseudo_1(\key, w||\counter)$, and generates a MAC on $\filter_c$ (together with a time stamp). If the server operates correctly, $\filter_c = \filter_s$ holds. Thus, only the MAC is uploaded to the server, which is then used by authorized users to check the integrity of the received $\filter_s$ to assure the correctness of the guessing counter value.
%authorized users can check the integrity of received $\filter_s$ via the MAC, which implicitly checks the correctness of the counter value.
%Note that the data owner does not need to upload $\filter_c$ because $\filter_c = \filter_s$ if the server operates correctly. In addition, $\filter_s$ can be used to verify the case that the queried keyword does not exist in the outsourced files. 

%hence the authorized users can leverage the counter value to check whether the server returned the correct number of file identifiers. In order to assure the authorized user obtaining the latest counter value, the data owner also generates the Bloom filter by adding $(\pseudo(\key, w||\counter)$ (which result in the same boom filter as that in the server), and signs the Bloom filter (together with a time stamp) with a secure signature scheme, where the signature will be stored in the server.  Therefore, the signature guarantees the freshness of the Bloom filter.

However, only assuring correct number of file identifiers is not enough, authorized users need to verify the correctness of the retrieved files with respect to the keyword $w$. To achieve this, each keyword is associated to an aggregate MAC, which is the result of aggregating MACs of all outsourced encrypted files containing $w$.

%the data owner generates an aggregate MAC for each keyword. For example, the files containing the keyword $w$ are generated in a sequence such as $\textbf{c}_w=\{c_1, \cdots, c_{cnt}\}$ in encrypted form. Note that the subscript of $c$ is not the file identifier, which is only used to represent the relative order of files containing keyword $w$, so $cnt$ is the latest counter for $w$. The latest MAC for keyword $w$ with counter $cnt$ is
%\setlength{\abovedisplayskip}{0pt}
%\setlength{\belowdisplayskip}{3pt}
%\begin{align*}
%\mu_{w||cnt} &= MAC_{k_s}(w||c_{cnt}) \oplus \mu_{w||cnt-1} \\
%&=MAC_{k_s}(w||c_1) \oplus\cdots\oplus MAC_{k_s}(w||c_{cnt}).
%\end{align*}
%Different from the basic construction, the data owner sets the latest entry for $w$ as:
%\begin{align*}
%$\langle \Phi_{k_s}(w || cnt_{prv})||key_{cnt_{prv}}||\mu_{w||cnt} \rangle\oplus \Phi_{key_{cnt}}(\Phi_{k_s}(w || cnt))$,
%\end{align*}
%and uploads the add token containing the MACs to the cloud. When the cloud replies a query for keyword $w$, it returns the search results, say, $c_1, \cdots, c_{cnt}$, and $\mu_{w||cnt}$ to the user. The user can verify the correctness of the search results by comparing whether $MAC_{k_s}(w||c_1) \oplus\cdots\oplus MAC_{k_s}(w||c_{cnt}) \stackrel{?}{=}\mu_{w||cnt}$.

\subsection{Main Construction}
Based on the above ideas, we present the full-fledged $\dsse$ construction as shown in Fig.~\ref{full-construction}, which highlights the difference from the basic construction in red color. The data owner maintains the state information (i.e., tuples of $(w, \counter, \gamma_{\counter})$) with a hash table $\hashtable_c$ mapping $w$ to $\counter, \gamma_{\counter}$, where $\gamma_{\counter}$ is the aggregation of the MAC for the concatenation of the file and $w$ so far.  The reason of concatenating the file and $w$ as input, rather using the file itself, is to prevent the replacement attack: given keyword $w_1$, the server might intentionally return the search result for another keyword $w_2$, an aggregate MAC and the set of file identifiers, which has the same number of file identifiers as that for keyword $w_1$.

Also, the data owner uses the timestamp $T$ (together with the Bloom filter $\filter_c$) to generate the MAC for preventing the replaying attack that the server might possibly return stale search result. We implicitly leverage the fact that the new file is periodically uploaded (e.g., every 10 minutes), so that 
authorized users can use the timestamp $T$ to assure the aggregate MAC is newly generated by the data owner.

Due to the lack of knowledge about $\counter$, authorized users (other than the data owner) generate the search token as shown in Fig.~\ref{authorized-user-query-construction}, where the WHILE loop is to guess  the counter value. Note that with the guessing counter value and the shared key from the data owner, authorized users are able to run $\sseverify$ to verify the correctness of the research result.
\vspace{-0.2cm}
\begin{figure}[htp]
\removelatexerror
\begin{center}
\fbox{
	\begin{tabular}{p{0.42\textwidth}}
	
$\underline{\token \leftarrow \gentoken(\keyset, \filter_s, r, w)}$:  After fetching the Bloom filter $\filter_s$ from the server, the authorized data user generates the search token as follows:

\begin{algorithm}[H]
Let $\counter$ = 1; \linebreak
\While{TRUE}{
  $\tau_\counter = \pseudo_1(\key, w||\counter)$ \linebreak
  \eIf{$\filterverify(\filter_s, \tau_{\counter})$ outputs 1} { 
     $\counter = \counter+1$
  }
  {
  $\counter = \counter-1$ \linebreak
  break;}
}
Compute $\key_{\counter} = \pseudo_1(\key, \hash(w||\counter))$; \linebreak
Let $\token = \se.\enc (r, \pseudo_1(\key, w||\counter)||\key_{\counter})$, which will be sent to the server.

\end{algorithm}
		\end{tabular}
	}
\end{center}
\caption{The algorithm for the authorized user generating search token. The data owner has already distributed $K =(\key, \key_{\se}, \key_{\mac})$ and $r$ to the authorized user.}
\label{authorized-user-query-construction}
%\vspace{-0.4cm}
\end{figure}

\noindent{\bf Optimization II: Speed up guessing the latest counter with binary search.} Instead of guessing the counter value linearly, authorized users can use the binary search to accelerate the guessing: The authorized user sets a large enough upper bound $Max$, and conducts the binary search for the latest counter $\counter$ within $[1, Max]$ such that $\pseudo_1(\key, w||\counter)$ is an element hashed to $\filter_s$ while $\pseudo_1(\key, w||\counter+1)$ not.

\noindent{\bf Optimization III: Reduce the number of elements hashed to $\filter_s$.} Note that the number of elements hashed into $\filter_s$ might become huge due to the increasing counter value $\counter$ when generating $\pseudo_1(\key, w||\counter)$ for keyword $w$. This results into a drawback: In order to keep low false positive rate, the size of $\filter_s$ becomes very large, which incurs costly bandwidth when authorized users retrieve it from the server. To get rid of it, the ``regular update'' strategy can be used: 
\begin{itemize}
\item Given the state information $\hashtable_c$, the data owner regularly (e.g., annually) generates a new Bloom filter $\filter_c$, which implicitly stores the current counter $\counter_{L}$ for each keyword $w$, generates the MAC and sends $\filter_c$ and the MAC to the server.
\item The server lets $\filter_s = \filter_c$ and  proceeds as in Fig.~\ref{full-construction}.
\item After receiving $\filter_s$, the authorized user extracts the counter $\counter_L$ first, and then guesses the latest counter starting from $\counter_L$.
\end{itemize}
By doing this, $\filter_s$ only contains elements with counters beginning with $\counter_L$ (rather than from $1$) for keyword $w$, and therefore its size can be reduced when keeping the same false positive rate. In addition, implicitly storing $\counter_L$ for keyword $w$ in $\filter_c$  can be done as follows: Given $\counter_L$, the data owner hashes $\pseudo_1(\key, w||\position||\digit_{\position})$ to $\filter_c$ where $\digit_{\position}$ is the least significant digit of $\counter_L$ when $\position = 1$, and the authorized user can guess $\counter_L$ by enumerating the combination of $\position=1, \ldots$ and $\digit_\position = 0,\ldots, 9$. For example, given $\counter_L= 456$ for keyword $w$, $\pseudo_1(\key, w||1||6), \pseudo_1(\key, w||2||5)$ and $\pseudo_1(\key, w||3||4)$ were hashed to $\filter_c$. Authorized users can guess $\counter_L$ by enumerating $\position$ and $\digit_\position$ and checking $\filter_s$ (since $\filter_s = \filter_c$) to determine whether $\pseudo_1(\key, w||\position||\digit_{\position})$ has been hashed into $\filter_s$ or not, until that there exists some $\position$ such that none of elements  $\pseudo_1(\key, w||\position||\digit_{\position}), \digit_{\position}=0, \dots, 9,$ was hashed into $\filter_s$.

%% file: security/security.tex
\section{Security Analysis}
\label{security}
We evaluate the security of our full-fledged construction to show that it achieves the security goals described in Section~\ref{design-goals}. We skip the formal proof here (which occurs in the full version of this work) due to the space limit. 

\noindent
\textbf{Data confidentiality}. The outsourced files are encrypted with the secure symmetric encryption together with secret key $\key_{\se}$. Without leaking $\key_{\se}$ to the server, data confidentiality is naturally assured by the secure symmetric encryption.

\noindent
\textbf{Index confidentiality}. Since each keyword in the index (i.e., $state_s$) is encrypted by the secure pseudorandom function $\pseudo_1$, without knowing the secret key, the server cannot learn the keyword from the index.

\noindent
\textbf{Forward privacy}. As discussed in the Section~\ref{sec:full-construction}, our construction encrypts the combination of the increasing counter and the keyword together, which makes the server unable to link the keyword in the newly added file to any stored encrypted keyword, without knowing the secret key $\key$. In addition, a secure pseudorandom function is used to mask the connection of tuples generated from the same keyword but with consecutive counter values, without knowing the corresponding secret key, the server cannot correlate these tuples together. That is, the server cannot know whether the newly added file contains any stored encrypted keyword, without knowing the secret keys for the pseudorandom functions. 

%In our scheme, if two subsequently added files $f_1$ and $f_2$ contain the same keyword $w$, the index entries for them will be $\mu_{k_s}(w||cnt)$ and $\mu_{k_s}(w||cnt^\prime)$ respectively, where $cnt^\prime=cnt+1$. As long as the pseudorandom function $\mu$ is secure and the secret key $k_s$ is kept confidential, the cloud cannot distinguish whether $\mu_{k_s}(w||cnt)$ and $\mu_{k_s}(w||cnt^\prime)$ are for the same keyword. Therefore, before searching for the keyword $w$, the cloud cannot know both files $f_1, f_2$ contain it, and thus forward privacy is achieved. Even if after search, some new files containing $w$, say $f_3, f_4$ are added, the forward privacy for them still remains until next search for $w$, since the new index entries are obfuscated by different $cnt$.

\noindent
\textbf{Search token privacy}. The keyword associated with the search token is protected with a secure pseudorandom function. Without knowing the key $\key$, the server cannot learn the keyword.

%The cloud cannot gain an advantage from the search token to infer the plaintext query, as long as RPF $\mu$ and $k_s$ are secure. Nevertheless, the cloud will know two search tokens are for the same keyword (i.e., search pattern), since $\mu$ is a deterministic function. All SSE schemes sacrifice search pattern for better search efficiency, in contrast, it can be achieved by public-key-based searchable encryption schemes at the cost of very low efficiency. 

\noindent
\textbf{Search capability enforcement}. Our construction implicitly shares the state information using the Bloom filter, and uses the group key to assure that only authorized users can generate valid search tokens. Therefore, the data owner can enforce the search capability securely (note that the cloud and users are not allowed to collude in our assumption).

%Since a successful query needs the cooperation between the cloud and the data user, any user who does not have the latest system parameter $r$ cannot cancel out the mask on the Bloom filter and submit a valid query to the cloud. As long as $r$ is kept confidential from the revoked users, the user revocation will not be violated.  

\noindent
\textbf{Verifiability}. Our construction uses the timestamp and the MAC to assure the freshness and  correctness of Bloom filter $\filter_s$, which further assures the correctness of the counter value for any keyword, forcing the server honestly returning correct number of the encrypted files. Moreover, the construction uses the aggregate MAC to assure the integrity of the returned files with respect to the keyword. Therefore, given the secure message authentication scheme, our construction assures that the authorized users and data owner can correctly verify the returned search result with an overwhelming probability.

%The completeness verification of our scheme relies on the verification on the Bloom filter, which is a bit array. If the cloud faithfully performs insertions to the remote Bloom filter $RBF$, the bit array of $RBF$ generated by the cloud will be identical to $LBF$ generated by data owner, and thus the MACs computed over $RBF$ or $LBF$ will be same as well. Without knowing the key $k_s$ used to generate $MAC_{k_s}(LBF||T)$ by the owner, the cloud cannot generate a different bit array $RBF'$ with non-negligible probability such that $MAC_{k_s}(RBF'||T) = MAC_{k_s}(LBF||T)$. The correctness verification depends on aggregate MACs. As proved in~\cite{katz2008aggregate}, as long as the cloud does not know $k_s$, it cannot forge a valid aggregate MAC over another set of $cnt$ faked files.

%% file: performance/performance.tex
\section{Performance Evaluation}
\label{performance}

In this section, we present the empirical performance result by simulating the e-healthcare system with the full-fledged $\dsse$ implementation.

\noindent{\bf Implementation:} We implemented the full-fledged $\dsse$ in JAVA, and instantiated $\pseudo_1, \mac$ with HMAC-SHA-1, $\pseudo_3$ with HMAC-SHA-512, $\se$ with AES and $\hash$ with SHA-1. In addition, we implemented all optimizations as mentioned above. We simulated the e-healthcare system by developing three separate processes for the data owner, the server and the authorized user respectively. The three processes communicate with each other via RESTful API, and were running in a laptop with 2.5GHz Intel i5 CPU, 8GB RAM and MAC OS.

\noindent{\bf Dataset:} 
In the experiment each PHI file consists of 15 pairs of attribute\footnote{The attributes include heartbeat, blood sugar, blood pressure, temperature and so on as in \url{http://www.clouddx.com/downloads/Heart-Friendly-Report-2015-12-24-092313.pdf} } in the format of attribute:value, which is treated as one single keyword (e.g., $w = heartbeat:75$). To simulate the scenario that the IoT gateway assembles and  uploads a new PHI file in every 10 minutes and lasts for 20-year, 1,051,200 synthesized PHI files were uploaded. 

\noindent \textbf{Performance on the Data Owner.} The average time for the data owner running $\addfile$ is 190 milliseconds, and the size of hash table (i.e, $\hashtable_c$) is around 1.3MB after uploading one million PHI files. The data owner also maintains a Bloom filter (i.e., $\filter_c$) of around 5MB by setting the false positive rate as $2^{-30}$, and updates it every year (i.e., after adding  $144\times 365=52,560$ new files as in Optimization III). We note that if the Bloom filter can be updated more frequently (e.g., less than every year), the size of the Bloom filter can be further reduced. 

\begin{figure}[t]
	\begin{center}
		\includegraphics[width=8.8cm, height=4.6cm]{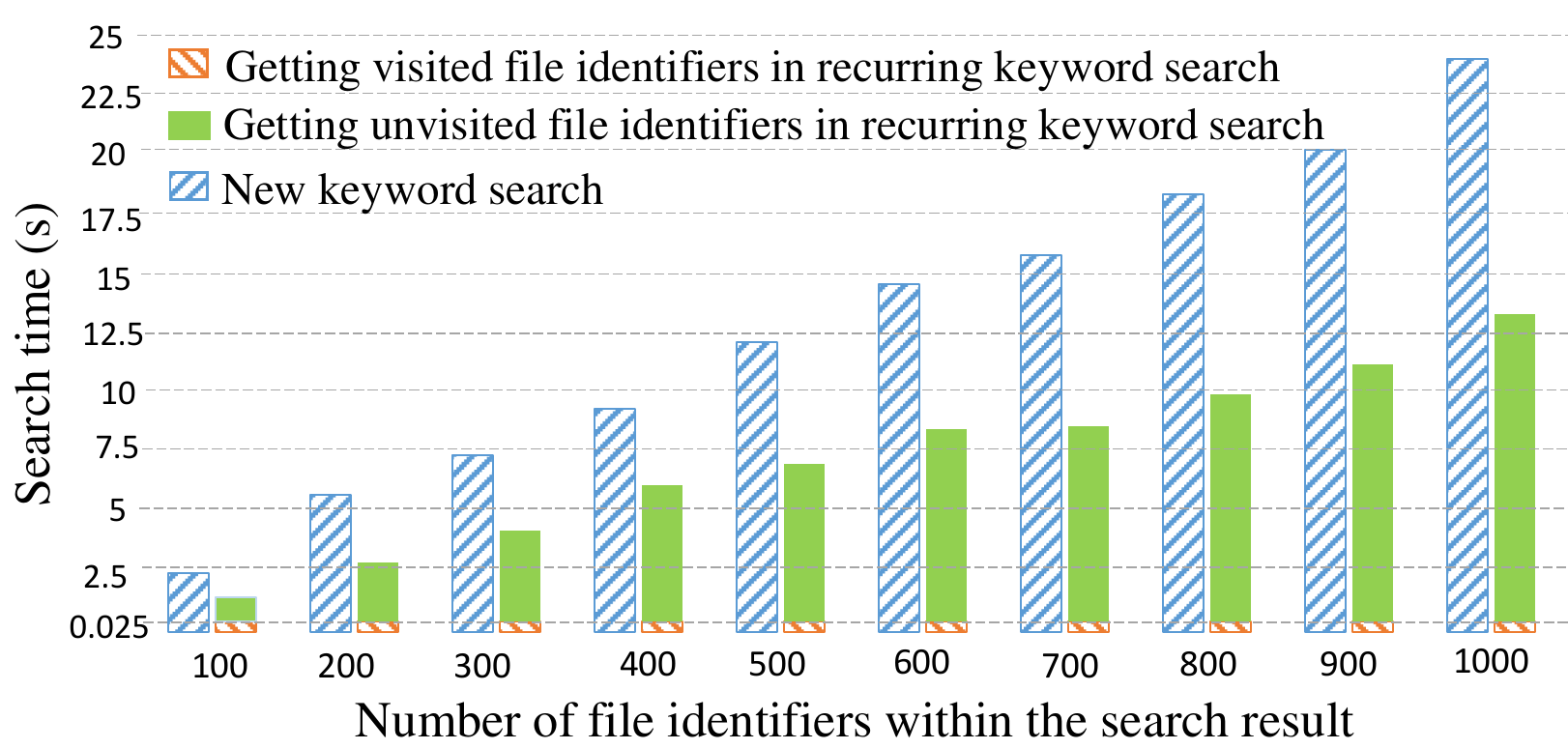}%[width=8.5cm, height=5cm]
	\end{center}
	\setlength{\abovecaptionskip}{-3pt}
	\caption{Performance for search operation running by the server storing one million files and the corresponding index (i.e., $\state_s$). Note that for recurring keyword search, half the number of file identifiers in the search result were newly added since the last time of the same keyword query (called unvisited identifiers), and the other half has been added to the server before the last time of the same keyword query (called visited identifiers). }%\protect\footnotemark
	\label{search_time}
	\vspace{-0.25cm}
\end{figure}

\noindent
\textbf{Performance on the Server.} We note that given a search token for keyword $w$ at time $T_{i+1}$, the  complexity of running $\search$ is linear to the number of encrypted files that contain $w$ and were uploaded within the interval of $T_{i}$ and $T_{i+1}$ (sublinear to the number of encrypted files), where $T_i$ is the last time when $w$ was searched for (The time of initializing the system can be regarded as $T_0$, at which the search result for any keyword is null). The reason is that, with Optimization I, the server stores in a consecutive manner all identifiers of encrypted files having $w$ at $T_{i}$, and can access them in a constant time thereafter. Therefore, we evaluated the search performance in the two scenarios: (i) new keyword search, which simulates that keyword $w$ has never been queried before, and (ii) recurring keyword search, which simulates that keyword $w$ has been queried before.  Fig.~\ref{search_time} shows the performance.  We can see that the search performance is closely related to the number of newly added files within the interval between two consecutive queries for the same keyword. In addition, we can see that the search performance is quite practical since returning $100$ files identifiers for new keyword search (resp. recurring keyword search) only costs around 2 seconds (resp. 1 second) (note that one million files and the corresponding index were stored in the server).

\begin{figure}[t]
	\begin{center}
		\includegraphics[width=8.8cm, height=4.6cm]{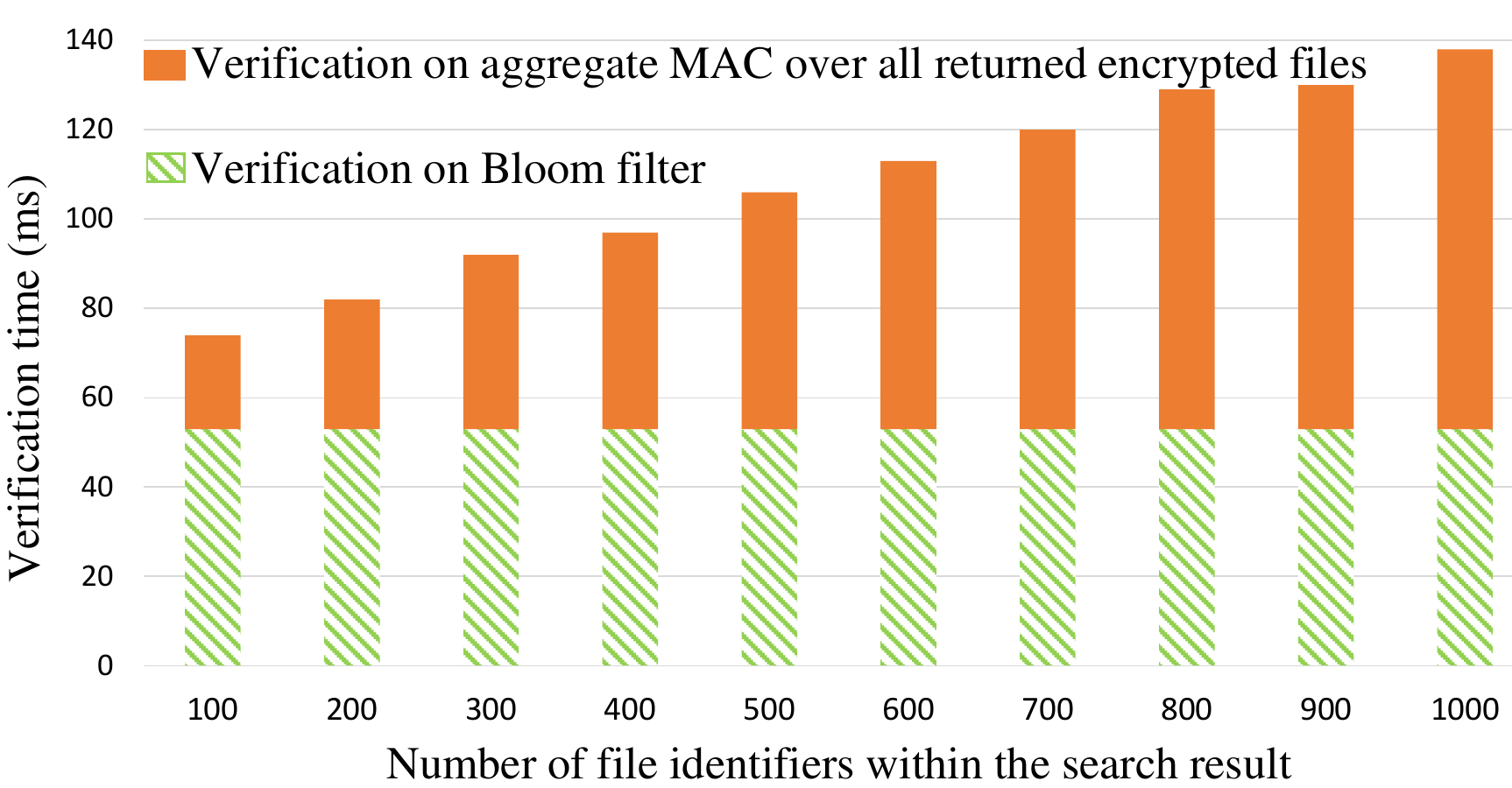}%[width=8.5cm, height=5cm]
	\end{center}
	\setlength{\abovecaptionskip}{-3pt}
	\caption{Performance for the authorized user verifying the correctness of the search result, i.e.,  verifying the correctness of the Bloom filter and aggregate MAC over all returned files.}%\protect\footnotemark
	\label{verification_time}
\vspace{-0.4cm}	
\end{figure}

\noindent
\textbf{Performance on the Authorized User.}
The time for the authorized user generating search token can be neglected (approximately 10 ms) due to the binary search (Optimization II). Therefore, we concentrated on the execution time for the authorized user verifying the correctness of the search result. The performance result is shown in Fig.~\ref{verification_time}, where we divided the verification time into two parts: one is for verifying the correctness of the Bloom filter (i.e., $\filter_s$ retrieved from the server) and the other one is for verifying the aggregate MAC over all returned files. We can see that the time for verifying the correctness of the Bloom filter is quite similar (e.g., around 55 ms in our experiments) no matter how many files are within the search result, and the time of verifying the aggregate MAC over all returned files is linear to the number of files. We can see that verification is practical because, even when dealing with the search result having 1,000 files, the verification time is only around 135 ms.

%% file: relatedwork/relatedwork.tex
\section{Related Work}
\label{relatedwork}

Cloud-assisted IoT system has become a popular design paradigm in many applications~\cite{tan2009ibe,li2011authorized,tong2014cloud,yang2016multi}, since the powerful computation and storage capabilities of cloud can overcome the constrains of IoT devices. This paper particularly relates to searchable encryption in e-healthcare:

\noindent
\textbf{Searchable Encryption}. Song \emph{et al.}~\cite{song2000practical} first explored the problem of searchable symmetric encryption and presented a scheme with linear search time. Curtmola \emph{et al.}~\cite{curtmola2006searchable} gave the first inverted index based scheme to achieve sub-linear search time. Although this scheme greatly boosts search efficiency, it does not support dynamic dataset. 
%which makes adding new files without leaking forward privacy very difficult. 
Since then, several schemes~\cite{kamara2012dynamic, kamara2013parallel, naveed2014dynamic, DBLP:conf/ndss/CashJJJKRS14, hahn2014searchable,stefanov2014practical} about dynamic SSE have been proposed, among which~\cite{kamara2012dynamic, kamara2013parallel, naveed2014dynamic, DBLP:conf/ndss/CashJJJKRS14} fail to provide forward privacy. Moreover, the previous work in~\cite{stefanov2014practical} offers forward privacy using a complicated hierarchical data structure, whereas the contribution in~\cite{hahn2014searchable} only achieves limited forward privacy (i.e., leaks the keywords contained in a new file if they have been searched for in the past). Besides dynamic SSE, verifiable SSE have been studied by ~\cite{kurosawa2013update,zheng2014vabks,cheng2015verifiable,sun2015catch}, which enables users to verify search results by using some verifiable structure such as the Merkle tree or an accumulator. However, previous work did not pay special attention to dataset with sequentially added files, which might leak additional information during the process of updating verifiable structure.

\noindent
\textbf{Secure data storage for e-healthcare}. Several searchable encryption schemes~\cite{tan2009ibe,li2011authorized,tong2014cloud} have been proposed for e-healthcare applications. Tan \emph{et al.}~\cite{tan2009ibe} proposed a lightweight IBE scheme to encrypt the sensing data and store it on a cloud. However, their public-key based scheme makes search over encrypted data very inefficient. Li \emph{et al.}~\cite{li2011authorized} presented an authorized search scheme over encrypted health data, which aims to realize search in a multi-user setting by enforcing fine-grained authorization before performing search operations. However, their search scheme is based on the predicate encryption, which is less efficient than SSE. Tong \emph{et al.}~\cite{tong2014cloud} proposed a SSE-based healthcare system, which achieves high search efficiency and partially hides the search and access patterns by using the redundancy. However, their scheme depends on a trusted private cloud and is not able to support dynamic data.

%% file: conclusion/conclusion.tex
\section{Conclusion}
%Considering the increasing adoption of cloud storage as well as various IoT initiatives in the healthcare industry, 
In this paper, we proposed a reliable, searchable and privacy-preserving e-healthcare system. The core of our system is a novel and full-fledged dynamic SSE scheme with forward privacy and delegated verifiability, which is dedicatedly designed to protect sensitive PHI files on cloud storage and enable HSPs to search on the encrypted PHI under the control of patients. The salient features such as forward privacy and delegated verifiability are achieved by a unique combination of the increasing counter, Bloom filter and aggregate MAC. Our experimental results and security analysis demonstrate that the proposed system provides a promising solution for meeting the stringent security and performance requirements of the healthcare industry in practice.           
\vspace{-0.1cm}

\label{conclusion}